# Highly Detailed Simulation for Connected Automated Vehicle Cooperative Driving

Andrei Fizulin
*HSE University*
Moscow, Russia
avfizulin@edu.hse.ru

Ilia Dolgov
*HSE University*
Moscow, Russia
imdolgov@edu.hse.ru

Andrei Karpukhin
*HSE University*
Moscow, Russia
avkarpukhin_1@edu.hse.ru

***Abstract*—End-to-end simulation of connected and automated vehicles requires consistent fidelity across mobility, environment modeling, V2X radio propagation, and decision-making/control modules. However, 2D representations of complex road infrastructure often fail to capture critical signal propagation dynamics, leading to overly optimistic connectivity assumptions. This paper presents a unified workflow within CAVISE that integrates map-based scene preparation with microscopic mobility simulation. The proposed framework incorporates a comparative analysis of trace-driven propagation in 3D versus a simplified 2D baseline, and a modular interface for integrating Autonomous Intersection Management (AIM) models. Collectively, these capabilities enable high-fidelity Cooperative Driving Automation (CDA) experiments. Leveraging ray tracing via Using Sionna RT at 5.9 GHz with the same radio and solver configuration for both geometry variants, we show that planar reduction in multi-level road infrastructure can remove physically present occlusions and substantially distort signal-loss dynamics. In a bridge overpass case study, the 2D baseline eliminates an occlusion interval observed in 3D, changing the outage behavior from intermittent to consistently connected and yielding an average signal-loss shift on the order of 10 dB. These propagation-induced biases highlight the inability of planar models to capture vertical occlusions necessitating 3D-aware communication modeling to ensure the validity of cooperative driving automation and intersection control evaluations.**

***Keywords*—connected and automated vehicles (CAV), cooperative driving automation (CDA), end-to-end simulation, radio propagation, ray-tracing, Sionna RT, Vehicle-to-Everything (V2X)***

## I. Introduction

The evolution of Intelligent Transportation Systems (ITS) is increasingly defined by the transition from autonomous to connected and cooperative systems. Cooperative Driving Automation (CDA) and cooperative perception [1] serve as the cornerstones of this next-generation ITS. However, the efficacy of these functions depends on timely Vehicle-to-Everything (V2X) message exchange under fast-changing propagation conditions, which require realistic perception and environment modeling to avoid qualitative distortions in simulation results [2]. Since large-scale field trials are expensive and difficult to reproduce, integrated simulation environments remain the primary means of validating cooperative functions prior to deployment. However, subsystem simplifications in sensing, environment representation, mobility, communications, or control may bias system-level conclusions and lead to misleading design decisions [3].

A persistent limitation in vehicular networking simulation is the continued reliance on planar (2D) propagation abstractions that cannot represent multi-level road infrastructure. Overpasses, underpasses, bridges, and stacked junctions create elevation-dependent occlusion and non-line-of-sight conditions that do not exist in a flattened scene. Prior work on three-dimensional (3D) vehicular networking has shown that these environments require explicit 3D consideration and has proposed methodologies to evaluate the resulting bias when the third dimension is ignored [4]–[7]. Tooling efforts such as Veins3D also reflect the practical need for 3D scenario support in V2X studies [8].

In parallel, CDA experiments increasingly incorporate Autonomous Intersection Management (AIM) logic that assigns priorities, produces control decisions, and orchestrates vehicle interactions at intersections. For such studies, the communication conditions are not merely a background component: burst losses, intermittent occlusion, and connectivity holes can alter the information available to the controller and thereby change qualitative outcomes. Consequently, reproducible CDA or AIM evaluation requires a workflow in which mobility, propagation, and control modules are coupled in a transparent and testable manner.

GPU-accelerated ray tracing has made high-fidelity propagation increasingly feasible. Ray-tracing engines such as OPAL provide physics-based modeling capabilities and have been investigated for end-to-end integration into CAV simulation stacks [9]. However, prior CAVISE-oriented evaluation of OPAL highlighted limitations of scenario-building workflows for representing true 3D environments, motivating approaches that operate directly on arbitrary 3D geometry [10]. More critically, a methodological gap remains in the application of these tools: existing studies often lack a controlled mechanism to isolate the specific impact of map verticality on cooperative decision-making, conflating errors from propagation model choice with errors from environment representation. NVIDIA Sionna RT offers such a unified approach: it supports GPU-accelerated ray tracing in general 3D scenes and enables controlled 2D and 3D scenario comparisons by applying geometry flattening while keeping the propagation engine and radio parameters unchanged [11], [12]. This property is particularly valuable for isolating the

This research was funded by the Russian Science Foundation (project No. 25-29-00551, https://rscf.ru/en/project/25-29-00551/).

effect of multi-level infrastructure on signal-loss dynamics without conflating different propagation model families.

This paper presents a comprehensive integrated workflow within the CAVISE. The framework combines: 1) precise map-based scene preparation; 2) microscopic mobility generation for realistic traffic dynamics; 3) physics-based propagation modeling computed for both full 3D environments and a flattened 2D baseline; and 4) integration architecture for Autonomous Intersection Management (AIM). Through this pipeline, we explicitly quantify how planar simplifications distort the communication signals required by downstream CDA models. This approach is demonstrated using a representative multi-level overpass scenario characterized by severe vertical occlusion and detail the interfaces that enable the systematic, reproducible benchmarking of diverse AIM strategies.

The remainder of this paper is organized as follows. Section II reviews related work on 3D vehicular propagation, ray-tracing-based modeling, and integrated CAV simulation. Section III describes the end-to-end workflow and experimental configuration, including the controlled 2D-versus-3D setup. Section IV reports propagation results for the multi-level overpass scenario and discusses their implications for cooperative applications. Section V presents the modular AIM integration architecture in CAVISE for CDA experiments. Section VI concludes the paper and outlines directions for future work.

## II. Related Work

Surveys of integrated simulation environments for connected automated vehicles emphasize that the credibility of end-to-end conclusions is bounded by the fidelity of the weakest subsystem, including communications [3]. For V2X propagation, multiple studies have highlighted that planar reduction can systematically fail in multi-level road environments and have provided methodology and measurement-based evidence motivating 3D modeling for V2X evaluation [4]–[7]. Practical tool support for 3D scenarios is also reflected in extensions such as Veins3D [8].

Geometry-based ray tracing has re-emerged as a feasible option due to GPU acceleration, enabling physics-based modeling of occlusion, reflections, and refraction in complex scenes [9], [11], [12]. This high-fidelity modeling is particularly relevant for CDA, where precise line-of-sight analysis determines the success of coordination and perception protocols. Integration efforts in the CAVISE context have shown both the value and the workflow challenges of using ray tracing in end-to-end simulation, motivating toolchains that can operate directly on general 3D geometry and support reproducible scenario building [10]. Recent work also explores integrating Sionna RT with network simulation to enable digital-twin style studies with realistic channels [18]–[21].

In contrast to studies that compare different propagation models or toolchains, this paper isolates the effect of geometry flattening by keeping the propagation engine and radio configuration identical between the 3D and 2D baselines.

## III. Workflow and Experiment Setup

This section details the end-to-end workflow and experimental configuration for generating comparable 3D and flattened 2D propagation traces in Sionna RT. The following subsections describe the pipeline overview, radio settings, baseline definition, and the overpass case study.

### A. Workflow Overview

The implemented workflow couples: 1) map-based 3D scene construction, 2) microscopic mobility generation, and 3) ray-tracing propagation evaluation. Road networks and surrounding objects are obtained from OpenStreetMap [13] and assembled as a 3D scene in Blender [14], using Blosm for OSM import where applicable [15]. SUMO generates vehicle trajectories as time-indexed poses (position, heading, speed) [16]. Sionna RT updates node poses and computes received power between transmitter–receiver pairs using ray tracing on the 3D geometry [11], [12]. In V2X simulation workflows, Artery is a commonly used middleware to couple traffic mobility and networking experiments [17]; the propagation traces produced here are intended for consumption by downstream networking and cooperative-application modules. The workflow is summarized in Fig. 1.

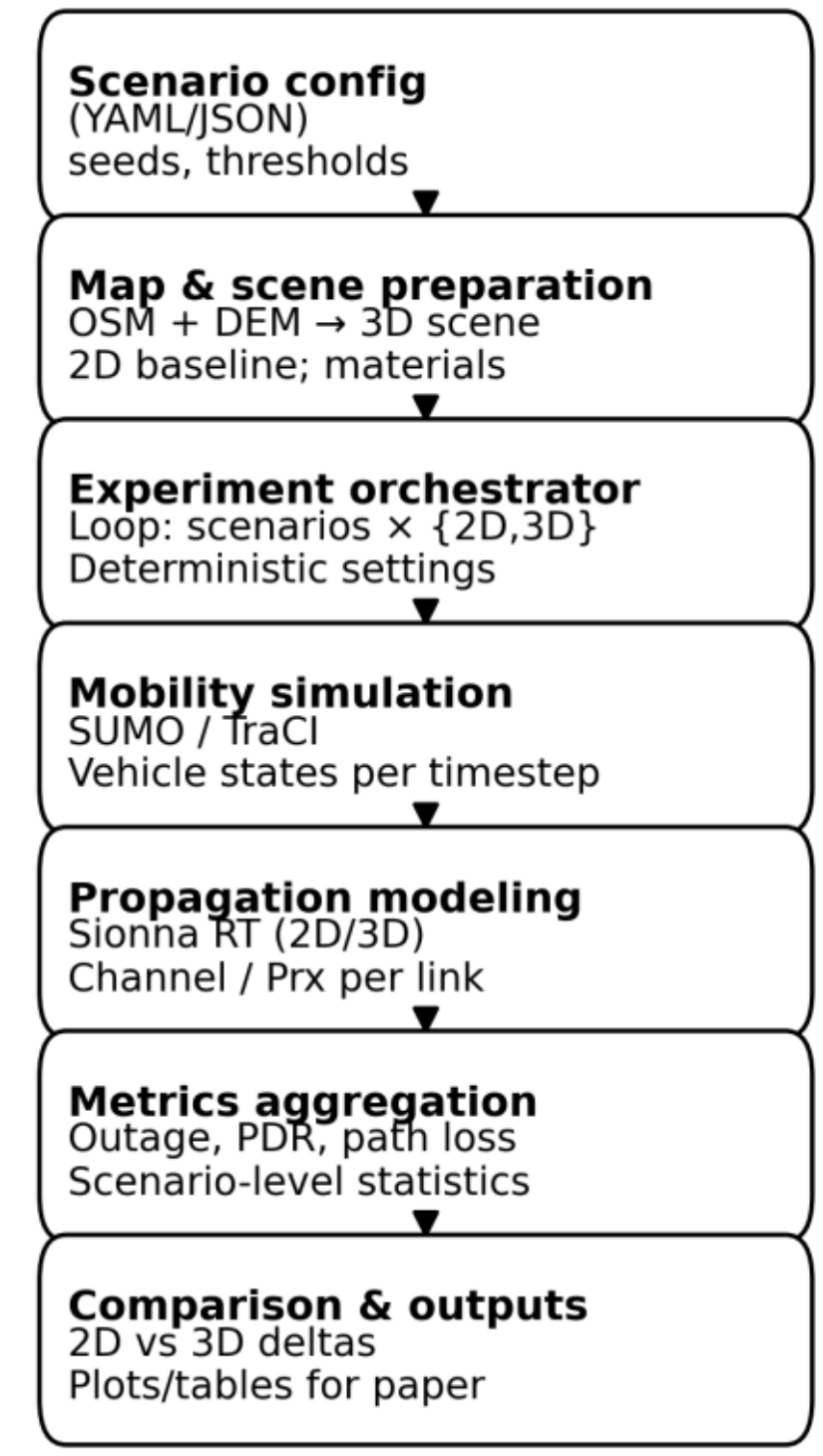

Fig. 1. Workflow overview: scene preparation, mobility generation, Sionna RT propagation evaluation (2D or 3D), and trace-based comparison.

### B. Propagation and Radio Configuration

We use NVIDIA Sionna RT (v1.0.1) as the ray-tracing engine in Path Solver mode. The 3D configuration and the flattened 2D baseline share identical radio parameters and solver settings to ensure a controlled comparison. The carrier frequency is 5.9 GHz and the transmit power is 20 dBm. The transmitter and receiver are configured as single-element isotropic antennas (1×1). We enable specular reflection, diffuse reflection, and refraction, and set the maximum path depth to 4 interactions. Signal loss is logged along the simulated trajectories at each channel update.

### C. Flattened 2D Baseline

To isolate the effect of elevation without changing the propagation engine, we prepare two scene variants: 1) a 3D variant with the original geometry, including elevation differences (e.g., bridge deck above an underpass), and 2) a flattened 2D variant obtained by collapsing vertical coordinates to a single plane while preserving horizontal footprints and object placement. All other settings are kept identical, making geometry flattening the only source of differences in the signal-loss traces.

### D. Case Study: Overpass Bridge Scenario

We select a road segment with an overpass and an underpass as a representative multi-level environment. Vehicles travel on the upper bridge and on the lower road beneath it. This class of infrastructure is explicitly highlighted in 3D vehicular network simulation literature as a scenario type where planar reduction changes physical visibility relationships [4]–[7]. In the 3D geometry, the bridge deck can occlude propagation between vehicles on different road levels when a receiver is under the deck. In the flattened 2D variant, the same occluding structure no longer exists as an overhead obstacle, which can lead to qualitatively different signal-loss dynamics. We therefore compare signal-loss traces for a representative cross-level link, where the receiver traverses the underpass while the transmitter remains on the upper road level.

## IV. Propagation Fidelity Results

This section reports the propagation fidelity results obtained from the controlled 3D versus flattened 2D comparison. We first analyze the signal-loss time series and outage behavior for the representative overpass link, and then discuss the implications of geometry-induced distortions for downstream cooperative applications.

### A. Signal-Loss Dynamics in 3D vs Flattened 2D

Fig. 2 and Fig. 3 compare signal-loss dynamics for the selected bridge link in the flattened 2D baseline and in the full 3D geometry, respectively. In the 3D case (Fig. 3(b)), the bridge introduces an occlusion interval that produces a pronounced increase in signal loss and intermittent outage. In the flattened 2D baseline (Fig. 2(b)), this occlusion disappears, resulting in a consistently connected link and substantially different temporal dynamics. In this case, flattening changes the outage rate from 42% (3D) to 0% (2D) and yields an average signal-loss shift on the order of 10 dB.

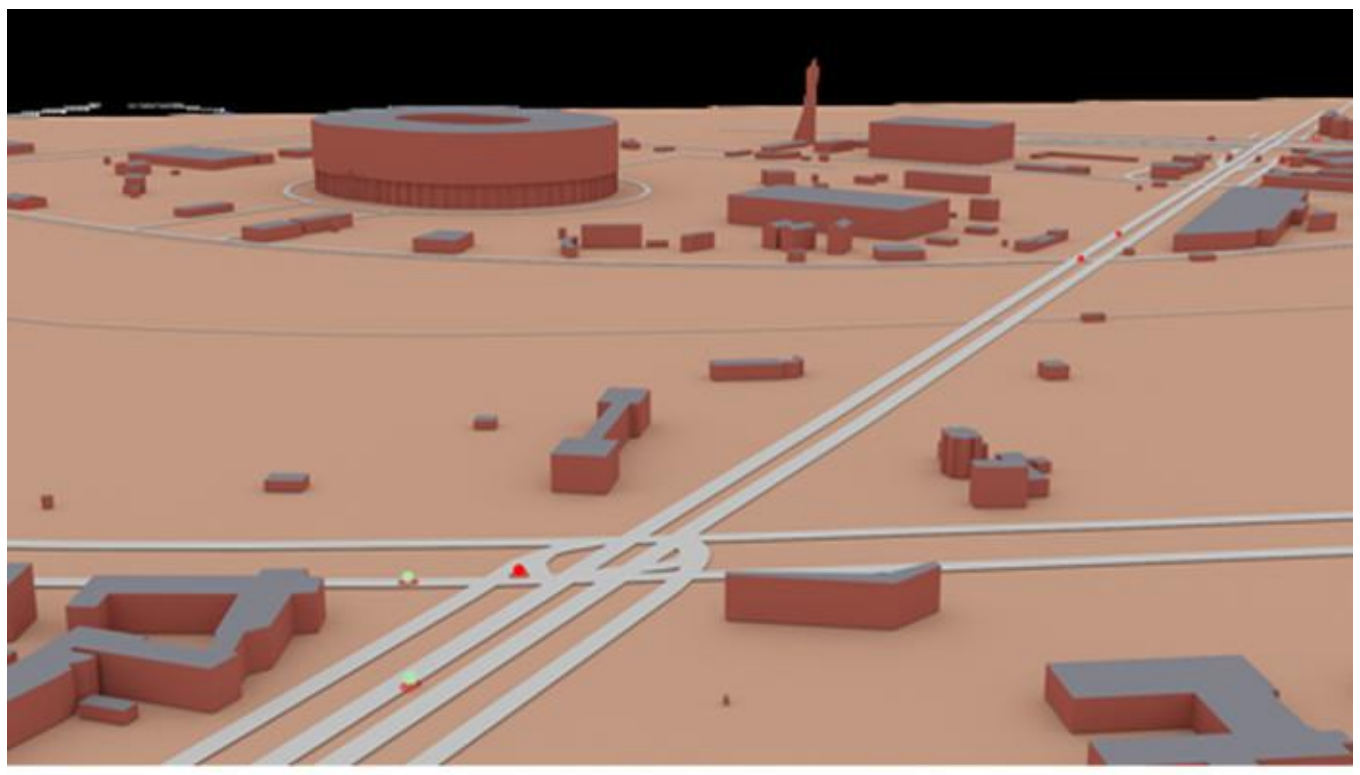

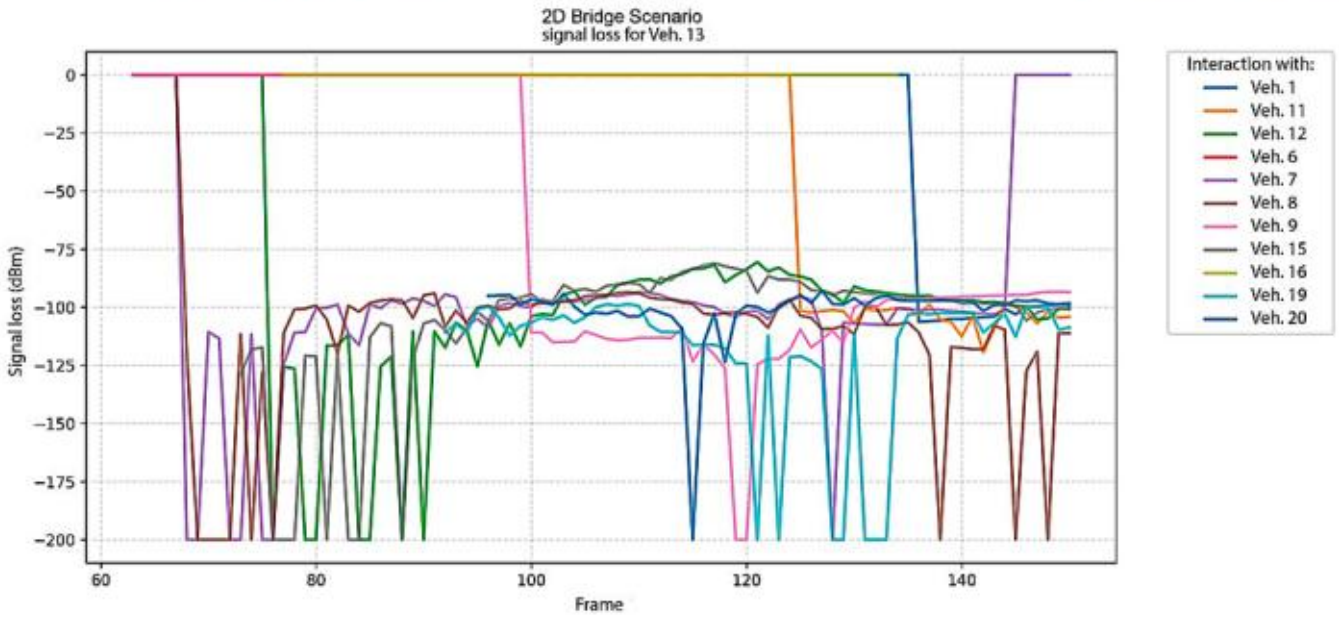

Fig. 2. Flattened 2D bridge baseline: (a) scene view (flattened 2D geometry); (b) signal loss (dBm) time series for Veh. 13 with multiple interacting vehicles.

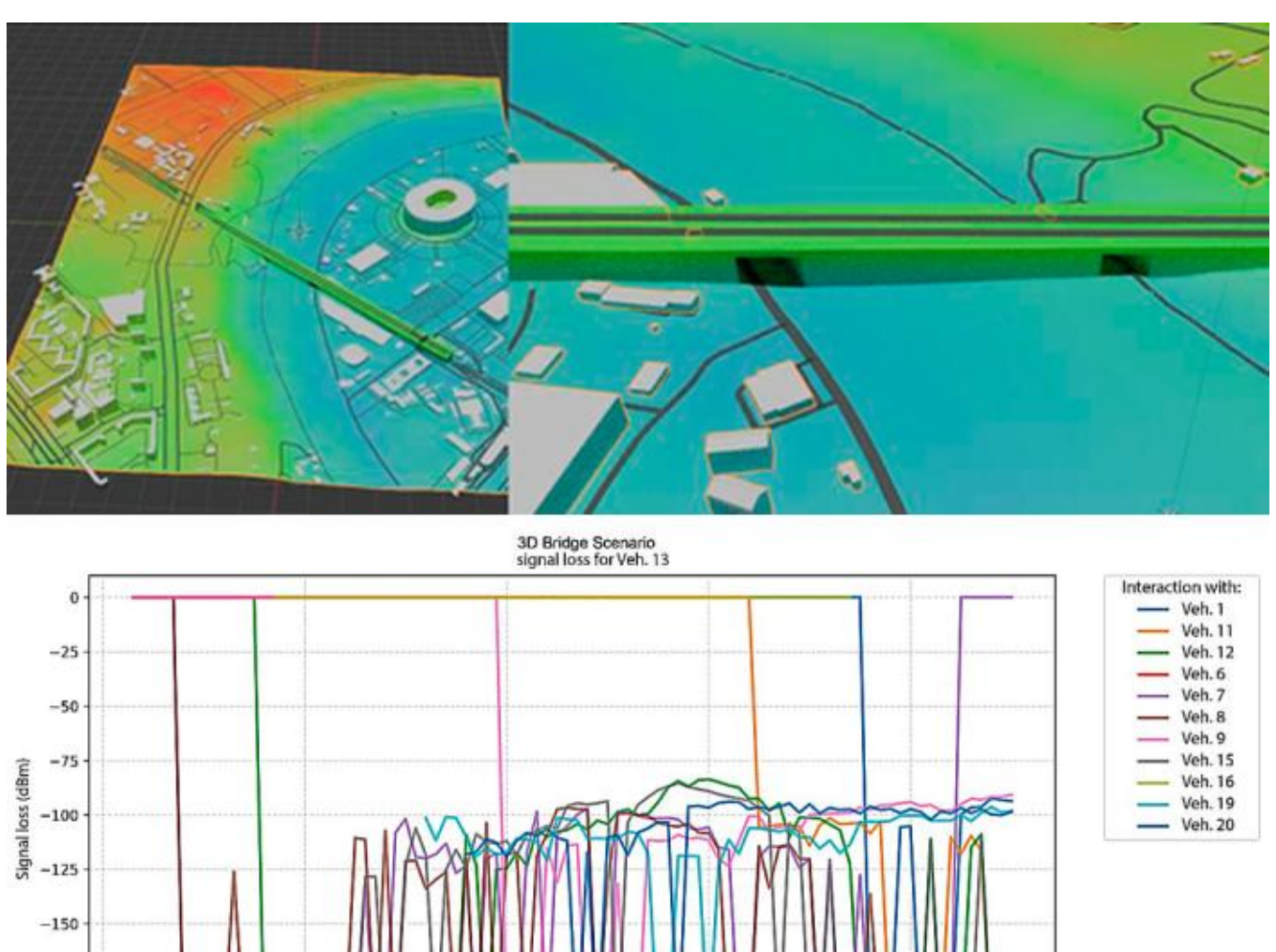

Fig. 3. 3D bridge geometry: (a) scene view (full 3D geometry); (b) signal loss (dBm) time series for Veh. 13 with multiple interacting vehicles.

### B. Implications for Cooperative Applications

Propagation-induced distortions of this type can translate into application-level bias. Cooperative driving and intersection management may appear more stable when burst losses and transient occlusions are removed by planar simplifications. The propagation traces produced in this work are intended to provide trace-consistent communication conditions for downstream CDA experiments in the end-to-end workflow.

### C. Reproducibility Notes

The presented comparison is reproducible given: 1) the OpenStreetMap region definition (bounding box and export date) [13]; 2) the Blender scene assembly settings, including OSM import configuration and object placement [14], [15]; 3) the SUMO network/routes configuration and the random seed used to generate mobility traces [16]; 4) the coordinate mapping between SUMO poses and the Sionna RT scene reference frame; and 5) the Sionna RT configuration reported in this paper. Sionna RT is used as the propagation engine [11], [12]. This artifact-based organization follows common practice in reproducible V2X digital-twin and ray-tracing-in-the-loop workflows [18]–[21].

## V. Experiment Limitations

The results in this paper are obtained from a controlled comparison between a full 3D scene and a flattened 2D baseline, where the ray-tracing engine, solver settings, and radio parameters are kept identical, and the only intended difference is the scene geometry flattening. The propagation evaluation uses NVIDIA Sionna RT (v1.0.1) in Path Solver mode with carrier frequency 5.9 GHz and transmit power 20 dBm. Transmitter and receiver vehicles are configured with single-element isotropic antennas (1×1). The ray-tracing configuration enables specular reflection, diffuse reflection, and refraction, and limits propagation paths by setting the maximum path depth to four interactions. The flattened 2D baseline is produced by collapsing vertical coordinates to a single plane while preserving horizontal footprints and object placement; all other settings are kept unchanged. The demonstrated comparison is based on the representative overpass bridge scenario described in the paper and on the selected bridge link used for the signal-loss time-series analysis. Therefore, the reported signal-loss dynamics and outage behavior correspond to this specific configuration and case study. Reproducibility depends on the availability and consistency of the inputs and assembly steps explicitly listed in the paper, including the OpenStreetMap region definition (bounding box and export date), the Blender scene assembly settings (including OSM import configuration and object placement), the SUMO network/routes configuration and the random seed, and the Sionna RT configuration used to generate the traces.

## VI. AIM Integration into CAVISE

In CAVISE, the propagation traces generated for multi-level infrastructure scenarios serve as the communication layer for downstream Cooperative Driving Automation (CDA) studies, including AIM-driven intersection decision-making [22]. These traces provide trace-consistent, time-varying link conditions (e.g., occlusion-driven burst losses and outages) that can materially affect perception sharing, coordination stability, and control safety margins. To enable systematic evaluation under such conditions, we developed a modular architecture for integrating and benchmarking AIM algorithms within the CAVISE workflow. The design supports controlled experimentation, repeatable model comparison across scenarios and seeds, and a clear path toward adapting validated AIM components for deployment. In this paper, the term AIM model refers to both machine learning–based models and classical algorithmic models.

### A. Architecture Overview

An AIM module is implemented as an independent software component with a well-defined interaction interface. This design enables integration not only with OpenCDA [23] but also with external systems (e.g., ROS). The modular approach ensures extensibility, reusability, and flexibility in the development and evaluation of intelligent mobility solutions. The overall integration architecture is shown in Fig. 4.

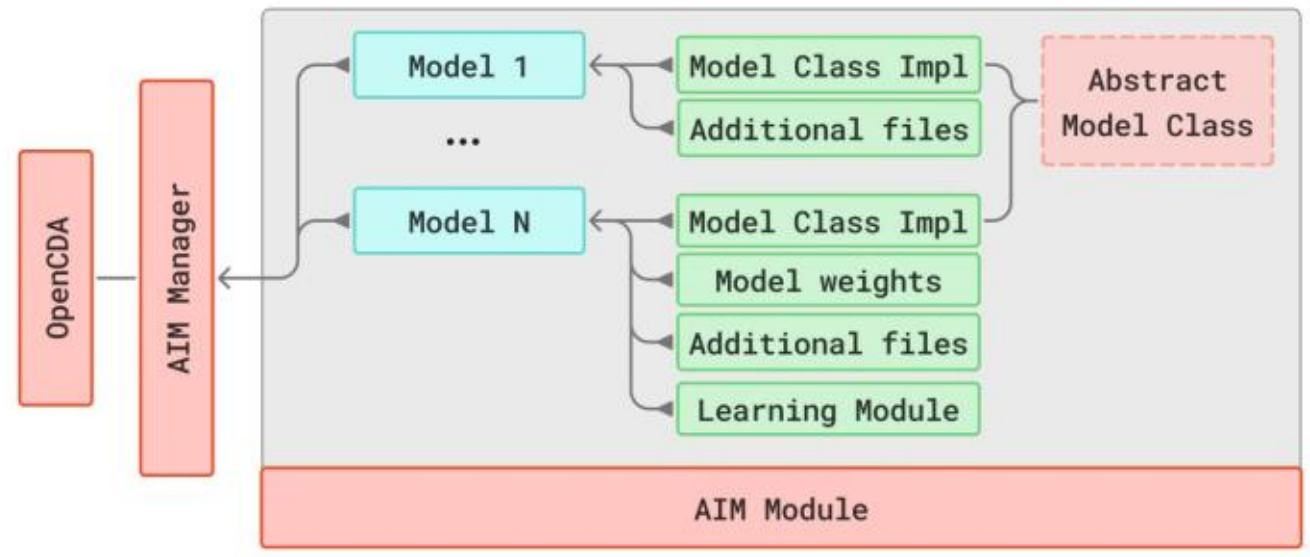

Fig. 4. AIM integration architecture in CAVISE (AIM Manager and pluggable AIM Module).

### B. AIM Manager

The AIM Manager serves as the central orchestration layer between the OpenCDA framework and various Autonomous Intersection Management (AIM) modules. This decoupled architecture enables a "plug-and-play" capability, allowing researchers to evaluate multiple decision-making and prioritization algorithms ranging from traditional heuristic-based scheduling to deep reinforcement learning models within a single experimental setup without modifying the core OpenCDA. The primary responsibilities of the AIM Manager include: loading and initializing AIM models; managing multiple AIM implementations (Model 1 – Model N); switching between different models while preserving a unified external interface; and receiving input data from OpenCDA and returning control outputs.

### C. AIM Module

An AIM Module represents an independent intersection control module. It supports multiple models, each of which consists of: an implementation of the model class; additional configuration files; optional training components and model weights, when learning-based approaches are used. This modular structure facilitates rapid prototyping, comparative evaluation, and systematic validation of different AIM models under identical simulation conditions. Fig. 5 illustrates an example of the CAVISE simulation environment during the execution of an AIM control scenario.

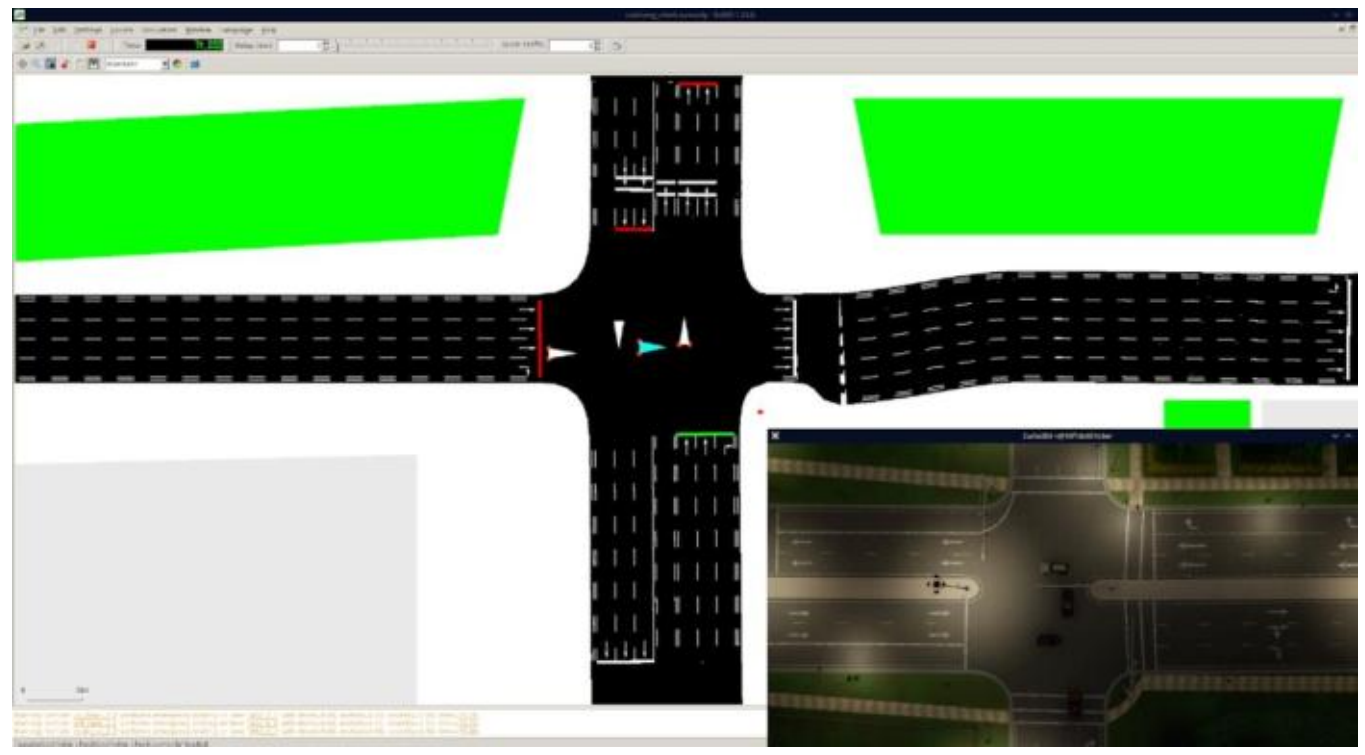

Fig. 5. AIM Scenario Demonstration

## VII. Conclusion

This paper combined two complementary components required for credible simulation-based evaluation of CDA technologies: propagation fidelity in multi-level infrastructure and AIM integration for experiments. We isolated the effect of geometry flattening by comparing 3D ray tracing and a flattened 2D baseline under identical propagation and radio configurations using Sionna RT. In the investigated overpass scenario, the 2D baseline produced deceptively optimistic connectivity, effectively removing physically present occlusion mechanisms. Quantitatively, the outage rate dropped from 42% in the 3D ground truth to 0% in the flattened baseline, accompanied by an average signal strength shift on the order of 10 dB. In parallel, we described an AIM integration architecture in CAVISE (AIM Manager and pluggable AIM Module) that enables systematic testing and switching among multiple AIM models through a unified interface without modifying OpenCDA. Together, these results motivate using 3D-aware communication modeling and trace-consistent experiment design when assessing intersection control and cooperative driving algorithms in simulation, since propagation artifacts can directly bias the operating conditions observed by downstream decision-making modules. Future work should focus on quantifying the impact of these 3D-induced propagation biases on the safety metrics and traffic throughput of specific AIM algorithms. Additionally, extending the unified CAVISE workflow to include dynamic 3D environmental changes, such as weather effects or seasonal foliage, remains a promising direction for enhancing simulation fidelity.